\documentclass[12pt,preprint]{aastex}
\usepackage{amsmath}
\usepackage{natbib}

\begin{document}

\bibliographystyle{apj}

\title{Constraining High Speed Winds in Exoplanet Atmospheres Through Observations of Anomalous Doppler Shifts During Transit}

\author{Eliza Miller-Ricci Kempton}

\affil{Department of Astronomy and Astrophysics, University of California, 
      Santa Cruz, CA 95064}

\email{ekempton@ucolick.org}

\author{Emily Rauscher}

\affil{Lunar and Planetary Laboratory, University of Arizona, 
  1629 East University Blvd., Tucson, AZ 85721}

\begin{abstract}

Three-dimensional (3-D) dynamical models of hot Jupiter atmospheres predict 
very strong wind speeds. For tidally locked hot Jupiters, winds at high 
altitude in the planet's atmosphere advect heat from the day side to the cooler
night side of the planet. Net wind speeds on the order of 1-10 km/s directed 
towards the night side of the planet are predicted at mbar pressures, which is
the approximate pressure level probed by transmission spectroscopy. These winds
should result in an observed blue shift of spectral lines in transmission on 
the order of the wind speed. Indeed, Snellen et al.~(2010) recently observed a 
$2 \pm 1$ km/s blue shift of CO transmission features for HD 209458b, which has
been interpreted as a detection of the day-to-night winds that have been 
predicted by 3-D atmospheric dynamics modeling. Here we present the results of 
a coupled 3-D atmospheric dynamics and transmission spectrum model, which 
predicts the Doppler-shifted spectrum of a hot Jupiter during transit resulting
from winds in the planet's atmosphere.  We explore four different models for
the hot Jupiter atmosphere using different prescriptions for atmospheric drag
via interaction with planetary magnetic fields.  We find that models with no 
magnetic drag produce net Doppler blueshifts in the transmission spectrum of 
$\sim2$ km/s and that lower Doppler shifts of $\sim$1 km/s are found for the 
higher drag cases, results consistent with --- but not yet strongly constrained
by --- the Snellen et al.~(2010) measurement.  We additionally explore the 
possibility of recovering the average terminator wind speed as a function of 
altitude by measuring Doppler shifts of individual spectral lines and spatially
resolving wind speeds across the leading and trailing terminators during 
ingress and egress.  

\end{abstract}

\keywords{planetary systems}

\section{Introduction \label{intro}}

Hot Jupiter exoplanets represent a fundamentally new class of planets that were
not anticipated and are not present within our solar system.  These planets
exist in extreme environments, residing very close to their host stars at 
semi-major axes of several hundredths of an A.U.  The mechanism that allows
hot Jupiters to migrate in to the near vicinity of their host stars from a 
formation location that was presumably much further out beyond the snow line
is still very much an open (and much debated) question.  In general, 
hot Jupiters fundamentally extend our understanding of planet atmosphere, 
structure, evolution, and migration into a new regime, and therefore further 
study of these planets aimed at characterizing and understanding their current 
states is warranted. 

In particular, hot Jupiters lie in a very interesting regime in terms of their 
atmospheric dynamics, in that they are expected to be tidally locked from 
simple timescale arguments \citep[e.g.][]{ras96}.  This leads to the planets 
having permanent hot day sides and colder night sides.  An important 
outstanding question is the extent to which heat is recirculated from the day 
side to the night side of these planets, which has important implications for 
their global energy budgets.  Some constraints have been made on day-to-night 
heat redistribution on hot Jupiters by observing the planets' IR emission as a 
function of orbital phase \citep{har06, knu07, cow07, knu09, cro10}.  However, 
the patterns seem to vary strongly from planet to planet.  More insight remains
to be gained from additional observations of this type along with direct 
constraints on the winds in hot Jupiter atmospheres, which ultimately govern 
the day-to-night heat flow.

Three-dimensional models of the atmospheric dynamics of hot Jupiters have been 
presented by a number of authors \citep[e.g.][]{sho09, dob10, rau10, hen11, 
thr11}, with the goal of understanding this new regime of atmospheric 
circulation.  While these models can vary considerably in the treatment and 
level of complexity of the underlying physics, a number of the qualitative 
results on hot Jupiter atmospheric dynamics have proved to be robust across 
most of the models.  These are (1) winds at pressures of $\sim 1$ bar set up an
equatorial jet that moves in the direction of the planet's rotation 
\citep[see][for an analytic description this behavior]{sho11}, (2) as a result 
of the equatorial wind pattern, the hottest point on the planet is shifted away
from the substellar point in the direction of the planet's rotation, and (3) 
winds higher in the atmosphere at $\sim$mbar pressures tend to flow directly 
from the hot day side to the cooler night side of the planet.  These three 
characteristics of the atmospheric dynamics models comprise a set of 
predictions for hot Jupiter wind and temperature patterns that can ultimately 
be tested through observations of the planetary atmospheres.  Already, 
observations constraining the brightness temperatures of certain hot Jupiters 
as a function of orbital phase have been found to be in general agreement with 
prediction (2) from the list above \citep{knu07, knu09}.  However, the observed
magnitude of the hot-spot shift from the substellar point varies to a greater 
extent than predicted by the circulation models \citep{cro10}.  Confirming the 
other two predictions -- (1) and (3) -- will ultimately require direct 
measurements of wind speeds in hot Jupiter atmospheres to confirm the nature of
the winds themselves.  

Recently \citet{sne10} produced the first tentative detection of exoplanet 
winds in the atmosphere of the transiting hot Jupiter HD 209458b.  Their 
measurement, if taken at face value, supports one of the theoretical model 
predictions of the wind flow pattern in hot Jupiter atmospheres -- that high
altitude winds should be directed towards the night side of the planet.  
\citet{sne10} obtained a high resolution transmission spectrum of HD 209458b, 
revealing the excess absorption of stellar light through the planet's 
atmosphere during transit.  At their high spectral resolution of $\sim$10$^5$ 
using the CRIRES spectrograph on the VLT, the authors were able to detect 
Doppler shifts in the planet's absorption lines during transit.  A linear drift
in velocity was detected throughout transit, which was attributed to the 
orbital velocity of the planet of $140 \pm 10$ km s$^{-1}$.  This direct 
detection of the planet's motion allowed for a direct measurement of the 
planet's mass, matching previous inferences of the planetary mass from stellar 
radial velocities (RVs).  Interestingly, when the Doppler signature of the 
orbital motion was subtracted off, \citet{sne10} found a remaining systematic 
$2 \pm 1$ km s$^{-1}$ blueshift in the planet's absorption signature.  This 
remaining blueshift has been attributed to the high-altitude winds in HD 
209458b's atmosphere, as predicted by theoretical atmospheric dynamics 
modeling.  These high-altitude winds should be directed towards the night side 
of the planet (and therefore directly towards the observer) at the $\sim$mbar 
pressure level probed by transmission spectroscopy, consistent with the 
observed net blueshift in the transmission spectrum.  In addition to the 
\citet{sne10} result, substantial blueshifts have now been detected in the the 
optical transmission spectra of several other transiting planets 
\citep{red08, jen11}, which might result from a combination of orbital motion
along with high-speed winds, but the implications of these observations have 
not been fully studied.

The interpretation of the \citet{sne10} result as a direct detection of 
high-altitude winds in the atmosphere of HD 209458b is compelling, but requires
further scrutiny.  Atmospheric dynamics models tend to produce wind speeds of 
$\sim$1-10 km s$^{-1}$ at mbar pressure levels \citep{sho09, dob10, rau10, 
hen11}, directed towards the night side of tidally-locked hot Jupiters.  
Interestingly, these wind speeds could be significantly reduced by magnetic 
effects at work in the planet's atmosphere.  The upper atmosphere is hot enough
to be weakly thermally ionized and, assuming a planetary magnetic field, the 
flow of the charged winds will induce a new component of the magnetic field 
with associated currents.  The winds will experience a drag -- \citet{per10a} 
show that this magnetic drag could potentially be strong enough to reduce zonal
wind speeds by up to a factor of $\sim$3 and alter the flow structure.  

Still, it is not straightforward to translate these wind speeds into the 
transmission spectral signature of a hot Jupiter atmosphere.  In this paper we 
attempt to do just that.  We couple together an existing 3-D atmospheric 
dynamics model \citep{rau10, per10a} with a transmission spectroscopy code 
\citep{mil09, mil10} such that the 3-D temperature-pressure and wind speed 
structures are self-consistently treated in the transmission radiative 
transfer.  The end result is to produce Doppler-shifted transmission spectra 
that include the effects of velocity shifts due to winds along with planetary 
rotation and orbital motion on the planet's absorption features.  Previous work
by \citet{spi07} studied the velocity effects of rotation on the transmission 
spectra of exoplanets but did not include the more dominant effect of winds.  
Our work also goes beyond previous work coupling 3-D atmospheric dynamics and 
radiative transfer models, which have self-consistently treated the 
temperature-pressure structure with the radiative transfer, while neglecting 
the Doppler shift inducing effect of winds \citep{for10, bur10}.

Our goal is to determine whether measurements of Doppler shifted winds in hot 
Jupiter atmospheres during transit could be used to constrain the nature of the
wind patterns themselves.  Furthermore, we attempt to determine whether a 2 
km~s$^{-1}$ blueshift, as seen in the \citet{sne10} transmission spectrum of HD
209458b, is consistent with the results of the atmospheric dynamics modeling.  
Specifically, we look at dynamics models that use different prescriptions for 
the treatment of drag in the atmospheres of hot Jupiters.  We study cases with 
no additional sources of atmospheric drag along with cases where prescriptions 
for magnetic drag have been included, to determine the overall effect on the 
planetary wind patterns as well as how these assumptions can alter the 
transmission signature of such a planet.  We also look into the possibility of 
recovering the spatial structure of exoplanet winds through transmission 
spectroscopy, both by using individual spectral lines to probe different 
heights in the planet's atmosphere and by measuring Doppler shifts during 
transit ingress and egress to separately map out wind speeds across the 
planet's eastern and western terminators.  

We present our model and methodology in Section~\ref{methods}.  Our 
transmission spectrum results are presented in Section~\ref{results}.  Finally,
we offer up some summarizing thoughts and conclusions in Section~\ref{concl}.  

\section{Methodology \label{methods}}

\subsection{3-D Dynamics Models \label{model}}

We present four models of hot Jupiter circulation here, two that are drag-free 
and two that include a simple treatment of magnetic drag in the atmosphere.  
The drag-free models differ in that one uses the same hyperdissipation strength
from \citet{rau10}, while the other uses a hyperdissipation timescale an order
of magnitude shorter in order to reduce the numerical noise that is apparent at
low pressures in the original model.  The two magnetic drag models differ in 
the implementation of drag at low pressures, as discussed below.  The 
three-dimensional models analyzed in this paper are all basic extensions of 
models presented previously in \citet{rau10} and \citet{per10a}, and details 
can be found in those papers.  Briefly, they were all calculated using the same
dynamical core and Newtonian relaxation scheme for radiative forcing as 
presented in \citet{men09}.  Each model was run for 2000 planet days at a 
resolution of T31L45 (corresponding to a horizontal resolution of 
$\sim 4^\circ$ and 45 vertical levels in log pressure).  Two modifications have
been made to the set-up of these models in order to better facilitate 
calculation of transmission spectra:  a different specific gas constant was 
used and the upper boundary has been extended up to 10 $\mu$bar.

The models from \citet{rau10} and \citet{per10a} used a specific gas constant 
of $R=4593$ J K$^{-1}$ kg$^{-1}$, which corresponds to a mean molecular mass of
1.81 g/mol, and a value for $R/c_p$ of 0.321 ($c_p$ is the specific heat 
capacity at constant pressure).  For better consistency with the transmission
spectroscopy modeling, which is highly sensitive to atmospheric scale height
and therefore mean molecular weight \citep[e.g.][]{mil09} , we chose to use 
$R=3523$ J K$^{-1}$ kg$^{-1}$ and $R/c_p =0.286$ in the circulation models 
presented here.  This now corresponds to a mean molecular weight of 2.36 g/mol,
a value more appropriate for solar composition, while the value for $R/c_p$ now
matches that for a purely diatomic gas.

The other significant change over the previously published models is that we 
have extended the upper boundary of the atmosphere to 10 $\mu$bar, necessary 
because transmission spectra probe pressures well above the 1 mbar top boundary
of the original models.  This extension to lower pressures meant that we had to
extrapolate our radiative forcing and drag prescriptions.

The Newtonian relaxation scheme used for the radiative forcing requires the 
choice of equilibrium temperatures and radiative times 
\citep[for details see][]{rau10}.  In order to extend these up to 10 $\mu$bar 
we chose to continue the 1000 K day-night temperature difference, which is 
constant down to 100 mbar.  The night side temperature was then set in the same
manner as in our previous work, so that at each level the integrated $T^4$ 
matched the profile from Figure 1 of \citet{iro05}.  The radiative times were 
taken from Figure 4 of \citet{iro05}, which includes these lower pressures.

\begin{figure}
\begin{center}
\plotone{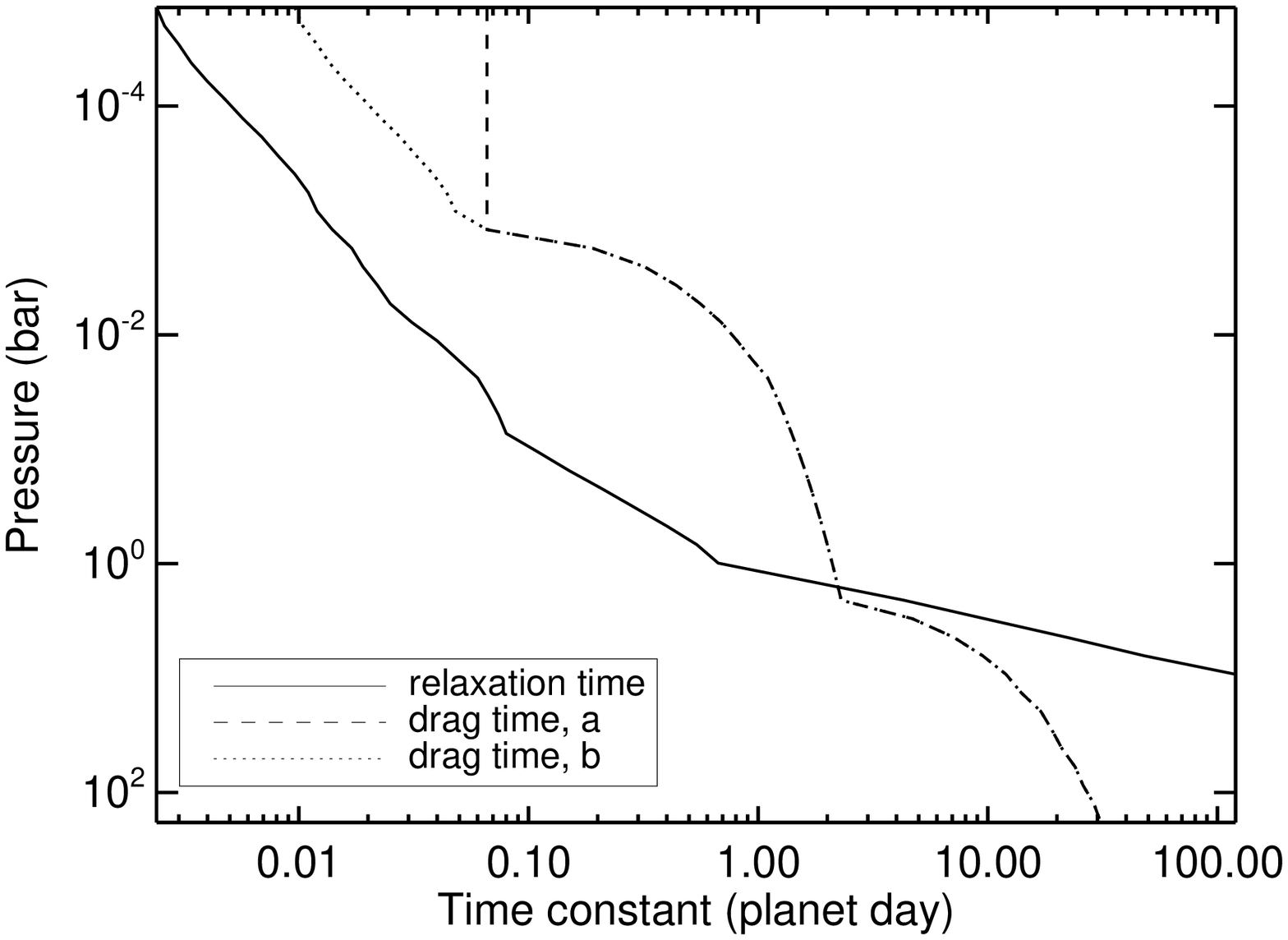}
\end{center}
\caption{Time constants used for the radiative forcing (\emph{solid}) and 
        magnetic drag schemes (\emph{dashed} and \emph{dotted}, two versions). 
	The times are given in units of a planetary day (= orbital period
	defined as 3.53 d).  The radiative times are set to be infinite a 
	pressures greater than 10 bar.} \label{fig:times}
\end{figure}

For a simplified treatment of magnetic drag in the atmospheres of hot Jupiters,
the models presented in \citet{per10a} employed a Rayleigh drag with a time 
constant that is horizontally uniform, but varies with pressure: 
$dv/dt = - v/\tau_{\mathrm{drag}}(P)$.  Here we use the strongest drag 
strengths from \citet{per10a}, which ranged from $2\times 10^4$ s at 1 mbar to 
$8\times 10^6$ s at 100 bar.  Since this scheme is already simplifying the 
underlying physics of magnetic drag, we choose two simple forms for the 
extrapolation of drag times to lower pressure: either we maintain a constant 
$\tau_{\mathrm{drag}}$ above 1 mbar, or we maintain a constant 
$\tau_{\mathrm{drag}}/\tau_{\mathrm{rad}}$ ratio above 1 mbar.  
Figure~\ref{fig:times} shows the radiative and drag times used throughout our 
models.

To summarize the four atmosphere models, they are:

\begin{enumerate}
\item{\emph{Drag-Free}: Canonical hot Jupiter model from \citet{rau10}, but extended to an upper boundary at 10 $\mu$bar and using an updated gas constant.}
\item{\emph{Drag-Free + Hyperdissipation}: Same as the drag-free model but with a hyperdissipation timescale that is an order of magnitude shorter to reduce numerical noise in the model.}
\item{\emph{Magnetic Drag (a)}: Magnetic drag model using the strongest drag strengths from \citet{per10a} below 1 mbar and maintaining a constant $\tau_{\mathrm{drag}}$ above 1 mbar.}
\item{\emph{Magnetic Drag (b)}: Same as magnetic drag model (a) but with a constant $\tau_{\mathrm{drag}}/\tau_{\mathrm{rad}}$ ratio above 1 mbar.  Since this means that the drag timescale continues to decrease with pressure, this model experiences stronger drag than model (a).}
\end{enumerate}

The atmospheric structure for all four models is qualitatively very similar to 
the results presented in \citet{rau10} and \citet{per10a}.  High in the 
atmosphere the basic flow pattern is from the hot day side to the cold night 
side, across all latitudes of the terminator, with transonic wind speeds 
(although weaker for the magnetic drag models).  At moderate pressures 
(hundreds of mbar) the advective timescales become comparable to the radiative 
timescales and the flow is able to alter the temperature structure from a 
hot-day/cold-night pattern to one where the hottest region of the atmosphere is
advected eastward of the substellar point.  Finally, at deep pressures the 
winds are much slower (and subsonic), but are easily able to minimize day-night
temperature differences.  Some details are different between the models 
analyzed here and those in \citet{rau10} and \citet{per10a}.  While similar, 
the advected temperature structures are not identical.  In addition, similar 
features between the models tend to occur at slightly lower pressure in the 
original models than the ones here, likely a result of the change in the 
specific gas constant and its effect on the pressure scale height.

\begin{figure}
\begin{center}
\includegraphics[scale=0.81]{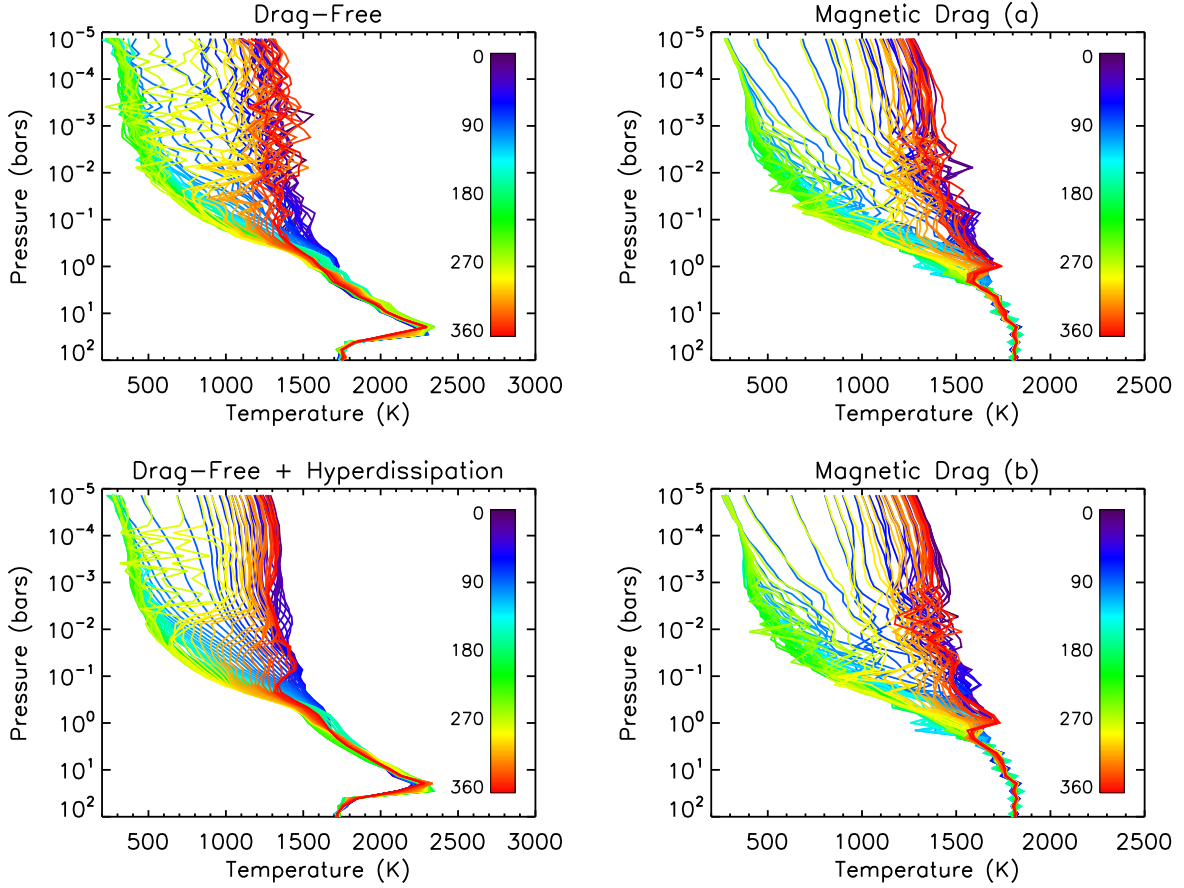}
\end{center}
\caption{Equatorial temperature-pressure profiles for each of the four hot 
        Jupiter models.  All profiles are taken at the equator for longitudes 
	from the substellar point ($\theta = 0^\circ$), eastward to the 
	antistellar point ($\theta = 180^\circ$), and back around to the day 
	side as indicated by the line color.  The eastward (trailing) limb 
	located at $\theta = 90^{\circ}$, is hotter than the westward limb for 
	all four models at moderate pressures, which is an effect of eastward 
	advection of hot dayside gas.
	\label{t_p}}
\end{figure}

The four models in this paper have generally similar temperature structures 
throughout the atmosphere, as shown by the temperature-pressure profiles in
Figure~\ref{t_p}.  Although the behavior at high pressure varies between the 
drag and drag-free models, all of the models have the same day-night 
temperature difference at low pressures (due to the same radiative forcing 
set-up and very short radiative times).  The models with magnetic drag or extra
hyperdissipation do not show the effect of small-scale numerical noise seen in 
the original drag-free model.  The temperature structure around the west 
terminator ($\theta=270^\circ$) is nearly the same as the east terminator 
($\theta=90^\circ$) for the magnetic drag models, but has a profile that has 
been more altered by advection in the drag-free models.

\begin{figure}
\begin{center}
\includegraphics[width=0.45\textwidth]{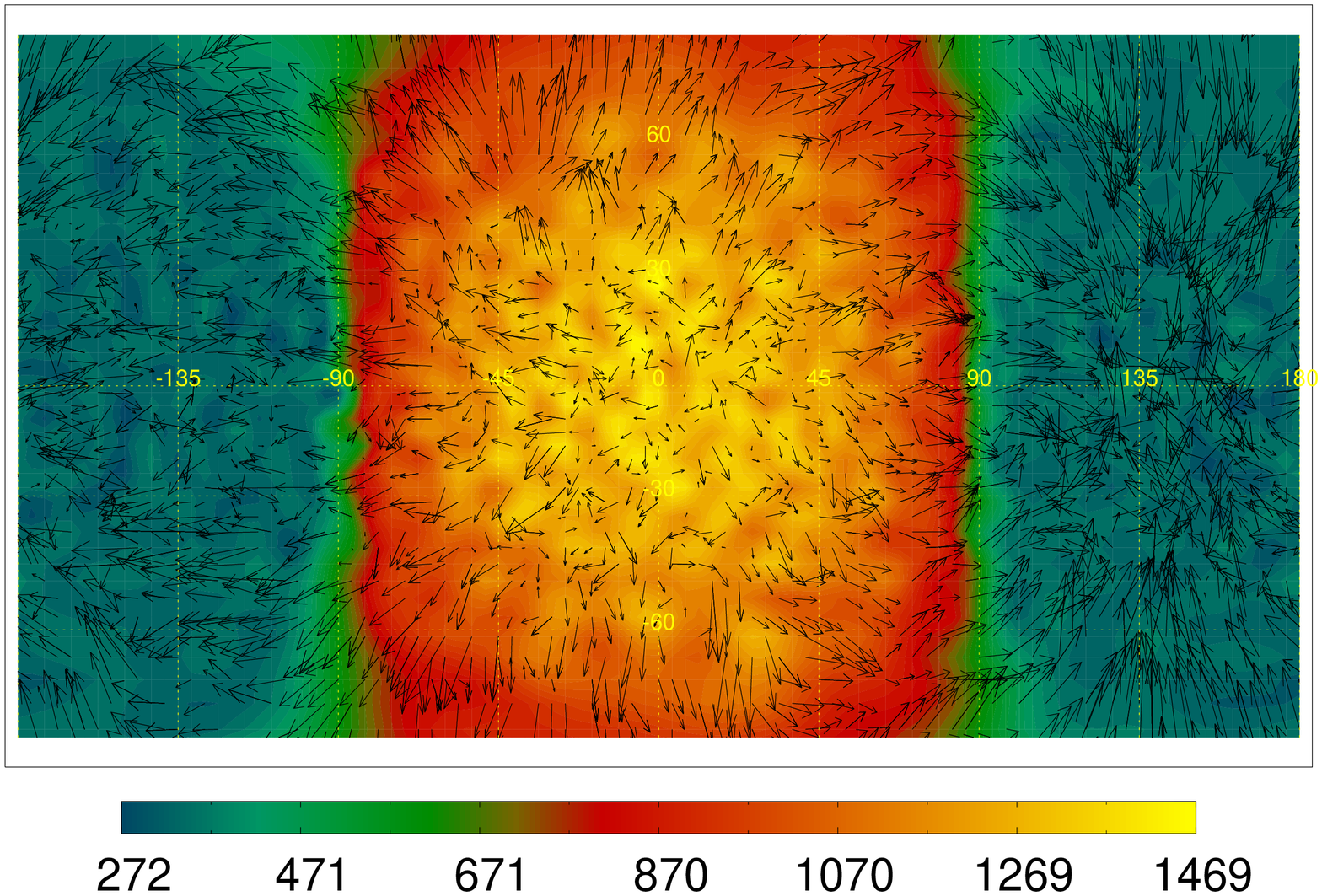}
\includegraphics[width=0.45\textwidth]{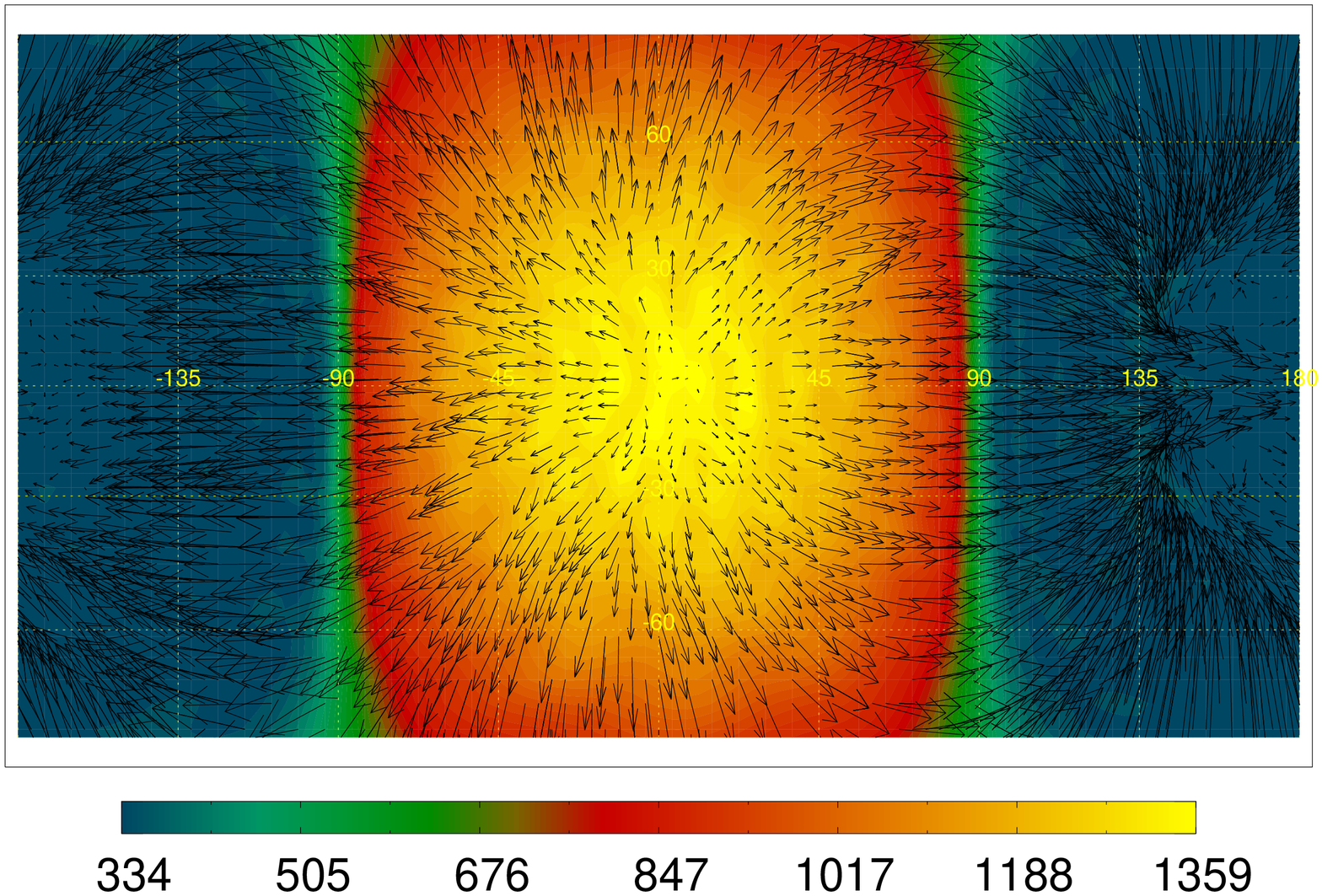}
\includegraphics[width=0.45\textwidth]{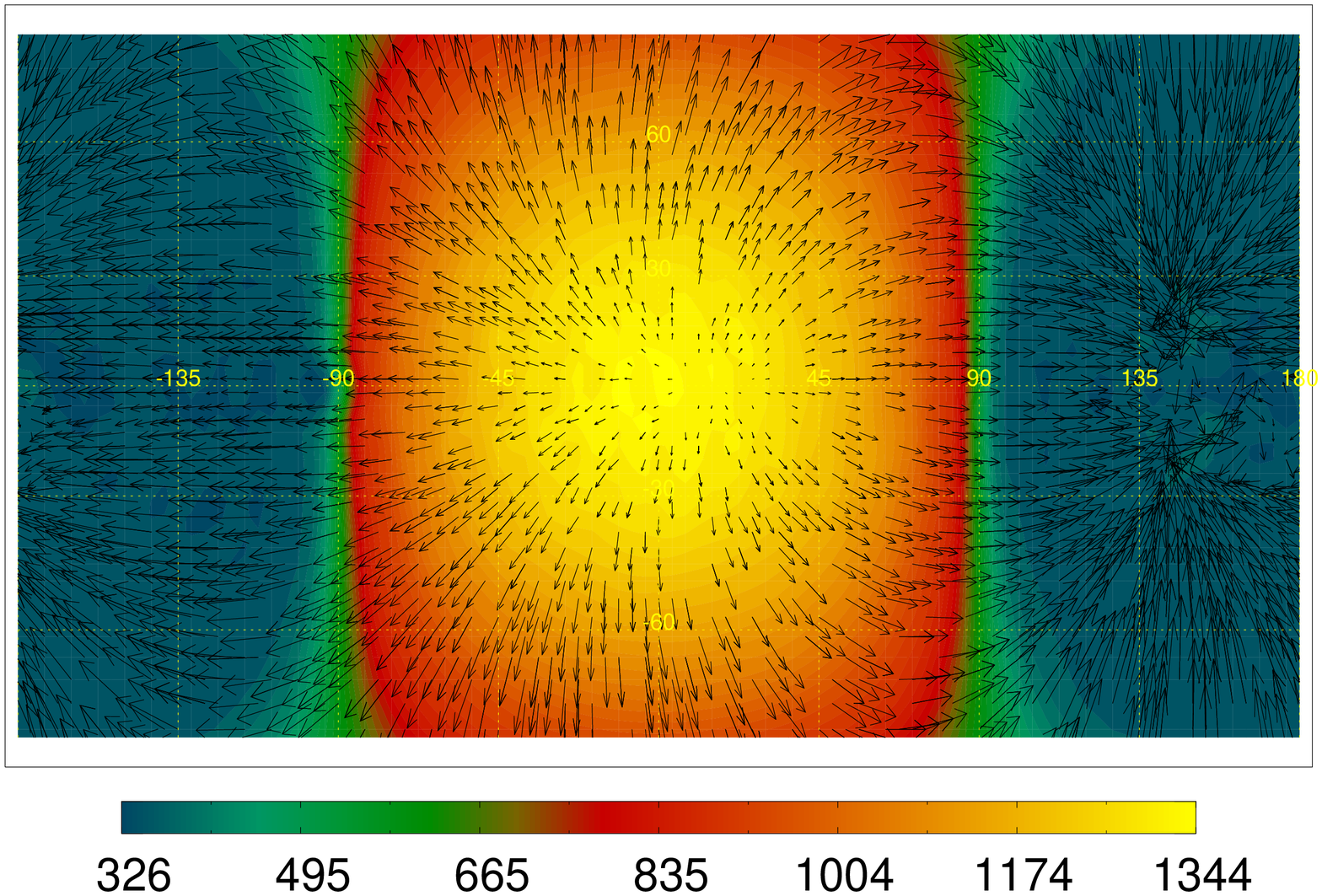}
\includegraphics[width=0.45\textwidth]{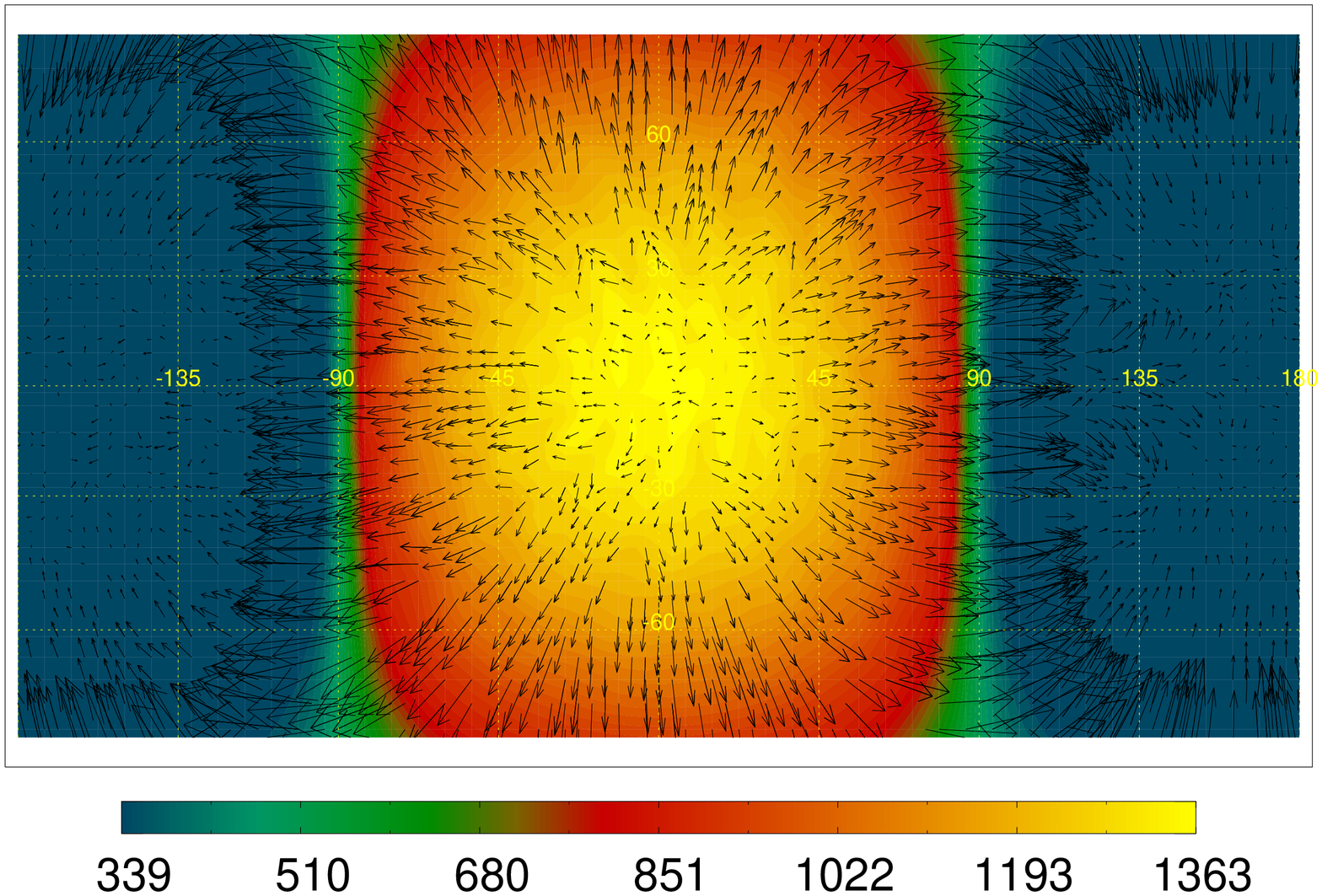}
\end{center}
\caption{Temperature map at the 60 $\mu$bar level, with two-dimensional wind 
        vectors overplotted.  (The center of the plot is the substellar point.)
	Shown are the original drag-free model (\emph{top left}), the drag-free
	model with enhanced hyperdissipation (\emph{bottom left}), and the 
	magnetic drag models, (a) which assumes $\tau_{\mathrm{drag}}$ constant
	above 1 mbar (\emph{top right}), and (b) with 
	$\tau_{\mathrm{drag}}/\tau_{\mathrm{rad}}$ constant 
	(\emph{bottom right}).  The maximum wind speeds in each of the models
        (clockwise from top left) are 15, 4.5, 2.5 , and 11 km/s.} 
        \label{fig:termu}
\end{figure}

The main difference between the four models is the flow structure, especially 
at the low pressures probed by transmission spectroscopy, where the Doppler 
effect could be observed.  The differences between the high-altitude flow in 
our four models can be seen in more detail in Figure~\ref{fig:termu}, which 
shows flow patterns across the planet at the 60 $\mu$bar level, representative 
of the high-altitude regime.  In all cases the winds are directed away from
the substellar point towards the anti-stellar point across the terminator.  
This results in a net blue-shifted wind directed towards the observer during 
transit for all models, although the magnitude of the wind speed and the details
of the wind pattern vary from model to model.  

First we compare the no-drag models.  Enhanced hyperdissipation in the second 
drag-free model reduces the numerical build-up of noise on small scales, which 
has the effect of making the flow more coherent, but reduces the maximum wind 
speeds.  Nevertheless, the mean eastward wind speed across the terminator is 
nearly the same for both models, and remains nearly identical between the two
models across a wide range of pressures ($\sim$5 km s$^{-1}$ at 60 $\mu$bar).  
The models with magnetic drag have much slower winds at this pressure.  The 
flow in model (b) is strongest in a narrow region on either side of the 
terminator, which is also the area probed by transmission spectroscopy, but it 
has an average eastward wind speed at the terminator of only $\sim$2.3 
km~s$^{-1}$, compared to $\sim$3.8 km~s$^{-1}$ in model (a).  The magnetic drag
models have similar average terminator winds throughout much of the atmosphere 
(where they have identical drag times), but at pressures less than 1 mbar the 
winds in the weak-drag model increase with altitude, while the winds in the 
strong-drag model decrease.

\subsection{Transmission Spectrum Radiative Transfer}

The transmission spectrum is obtained by dividing the spectrum obtained during
transit by the stellar spectrum, thus revealing the excess absorption from
the species that make up the planetary atmosphere.  We calculate the 
attenuation in intensity of a beam of stellar light passing through the 
planet's atmosphere according to 
\begin{equation}
I(\lambda) = I_{0} e^{-\tau}, 
\label{intensity}
\end{equation}
where $I_0$ is the incident intensity from the star.  We ignore any effects
of additional scattering into or out of the beam or refraction, which have 
been found to be of minimal affect to hot Jupiter transmission spectroscopy 
\citep{hub01} .  The slant optical depth $\tau$ is calculated as a function of 
wavelength $\lambda$, height $H$, and latitude $\phi$ according to
\begin{equation}
\tau = \int \kappa ds,
\label{tau}
\end{equation}
where $ds$ is the differential path length through the planet's atmosphere 
along the observer's line of sight, and $\kappa$ is the opacity calculated for
solar composition gas evaluated at the local temperature and pressure, which
are each in turn dependent on the local height $H$, latitude $\phi$, and
longitude $\theta$ according to the 3-D dynamics model.  The opacity is 
furthermore Doppler shifted by the local line-of-sight velocity as outlined in 
the following paragraphs.  We include opacities from gas phase CH$_4$, CO, 
CO$_2$, H$_2$O, and NH$_3$ along with collision-induced H$_2$ opacities.  
Further details on our opacity tables and equilibrium chemistry calculations 
for solar composition gas can be found in \citet{mil09} and \citet{mil10}.  It 
is important to note for this work that our opacity tables of the molecular 
absorption come in the form of line lists, which allows us to calculate 
opacities at arbitrary spectral resolution.  The total in-transit flux passing 
through the planet's atmosphere is then calculated by integrating the 
intensity from Equation~\ref{intensity} over the solid angle subtended by the 
atmosphere.  

As photons pass through the planetary atmosphere during transit, they encounter
gas moving at the local velocity $v$, whose line-of-site component will induce 
a Doppler shift on the absorption signature.  The Doppler shift through the 
planet's atmosphere is given by
\begin{equation}
\frac{\Delta \lambda}{\lambda} = \frac{v_{LOS}}{c},
\end{equation}
where $v_{LOS}$ is the line-of-sight component of the velocity.
There are three components that contribute to the line-of-sight velocity
velocity that a photon ``sees'' as it passes through the planet's atmosphere.  
These are (1) the wind speed, (2) the rotational velocity, and (3) the orbital 
speed, all evaluated locally in the planet's atmosphere.  The net line of sight
component to the velocity is then given by
\begin{equation}
v_{LOS} = -(u \sin \theta \cos \phi + v \cos \theta \sin \phi + (R_{p} + z) \Omega \sin \theta \cos \phi + v_{orb} \sin \varphi).
\label{vlos}
\end{equation}
The first two terms in Equation~\ref{vlos} give the contribution to the 
line-of-sight velocity from the winds in the planet's atmosphere.  The 
east-west and north-south components of the wind are given by $u$ and $v$, 
respectively.  The third term in the equation gives the contribution from the 
planet's rotation, where $R_p$ is the planet's radius at 1 bar 
($=1.32$ $R_{Jup}$ for all models), z is the height in the atmosphere above the
1-bar level, and $\Omega$ is the planet's rotational speed in radians~s$^{-1}$.
The final term in Equation~\ref{vlos} gives the contribution from the orbital 
motion of the planet.  The orbital speed is denoted by $v_{orb}$, and $\varphi$
is the phase angle of the orbit, defined as $\varphi = 0$ at the center of 
transit.  For the results presented in this paper we assume a circular orbit 
resulting in a constant $v_{orb}$.  To calculate the rotation and orbital 
speeds we assume a tidally locked planet ($P_{orb} = P_{rot}$) with an orbital 
period of 3.53 days.  Figure~\ref{fig:dshift} shows the Doppler-shifted wind 
speeds along the planet's terminator that result from each of the four 
atmospheric dynamic models described in Section~\ref{model}.  The opacity 
$\kappa$ from Equation~\ref{tau} is finally evaluated at wavelength
\begin{equation}
\lambda = \lambda_0\left(1-\frac{v_{LOS}}{c}\right),
\end{equation}
where $\lambda_0$ is the unshifted wavelength, to produce the properly 
Doppler-shifted absorption.  


High resolution spectroscopy at a spectral resolution of $\sim$10$^5$ along 
with sufficiently high signal-to-noise is necessary to measure Doppler shifts 
in transmission spectra at the km~s$^{-1}$ level.  Here we calculate all of our
transmission spectra at a spectral resolution of $10^6$, which is higher than 
the resolution of most currently available high-resolution spectrographs by at 
least an order of magnitude.  (For example, the CRIRES spectrograph used by 
\citet{sne10} has a working resolution of $\sim$10$^5$).  We compute spectra at
such high resolution to clearly show the effects of Doppler shifts on our 
transmission spectra and also to show the power of very high spectral 
resolution for future instrumentation.  All of our spectra can be easily 
degraded to lower spectral resolution by convolution with a Gaussian of the 
appropriate width.  

We calculate all of our spectra from 2291 to 2350 nm as a representative 
wavelength range over which high resolution transmission spectra can be 
obtained from the ground.  This wavelength range covers the 2.3 $\mu$m first 
overtone ($\Delta \nu = 2$) band of CO.  This is also the wavelength coverage 
of the observations by \citet{sne10}, which facilitates comparisons between our
model spectra and their results.

It is important to note that the stellar spectrum experiences velocity shifts 
of its own during transit.  These shifts result from both the induced motion 
from the planet's orbit (stellar radial velocity) along with the 
Rossiter-McLaughlin effect by which the planet blocks out a portion of the 
blue- or red-shifted limb of the star during transit.  Both effects produce
a zero net Doppler shift at the center of transit, provided that the planet's 
orbit is circular and the stellar spin axis is aligned with the normal axis of 
the planet's orbit.  However, when either of these effects becomes non-zero, it
is necessary to divide out the appropriately Doppler-shifted stellar spectrum
from the in-transit spectrum to obtain a transmission spectrum that only has
the effects of the planetary Doppler shift imprinted on it.  For the spectra
that we present in the following section, we assume that the stellar Doppler
shift is known and has been appropriately accounted for.  

\section{Results \label{results}}

\subsection{Transmission Spectrum Doppler Shifts}

The entire 2291-2350 nm section of the hot Jupiter transmission spectrum for 
the drag-free atmosphere model is shown in Figure~\ref{transmission_full} with 
no velocity shifts.  In the unshifted spectrum, the first overtone band of CO 
is clearly visible at the model spectral resolution of $10^6$.  Many additional
spectral features from water are also present in this region of the spectrum
along with weaker CH$_4$ features that originate from cooler regions in the 
atmosphere.  The comb of strong and well-separated CO lines around 2.3 microns
is particularly useful for measuring Doppler shifts as long as the spectral 
resolution exceeds $\sim$10$^4$.  At the much lower spectral resolution of most
transmission spectrum measurements to date, the 2.3 $\mu$m CO features
are unresolved in a single broad absorption band, and RV measurements at a 
precision of $\sim 1$ km~s$^{-1}$ are not possible.  

Figure~\ref{spectra_snapshot} shows a representative 2.3 nm segment of the 
fully Doppler shifted spectra from 2308.0 to 2310.3 nm.  (The rest of the 
spectrum shows similar RV shifts from winds, but it is not feasible to show the
entire spectral range of our model in a single plot and to still see the 
Doppler shifts by eye.)  A net blueshift in all four models is clearly apparent
in the top panel of Figure~\ref{spectra_snapshot}.  Additionally, a significant
amount of line broadening occurs in the transmission spectra from the joint 
effects of winds and rotation, which effectively weakens the peak strength of 
each of the individual spectral lines and also causes many of the adjacent 
lines to blend together.  The effect of broadening due to rotation alone is 
shown in the bottom panel of Figure~\ref{spectra_snapshot}, clearly indicating 
the extent to which the Doppler broadening is due to rotation as opposed to 
winds.  Rotation is the dominant source of broadening (quite similar to the 
effect of rotational broadening on stellar spectra).  However, the winds 
themselves also cause some further broadening of the transmission spectra, 
since the wind patterns are not entirely coherent and instead some considerable
variation in the line-of-sight wind speeds occurs around the terminator as 
shown in Figure~\ref{fig:dshift}.  Some of the models clearly show a higher 
level of incoherence in the wind pattern than others (e.g. the standard 
drag-free model as opposed to the model with enhanced hyperdissipation), which 
has a small effect on the width of the broadened spectral lines from model to 
model.

By cross correlating the Doppler-shifted spectra against the unshifted 
transmission  spectra, we can obtain the average velocity shift for the entire
2291-2350 nm region of the calculated spectrum.  The cross correlation 
functions for each of the four atmosphere models taken at the center of transit
(phase $\varphi = 0$) are shown in Figure~\ref{crscor}.  Each model gives a
different overall velocity shift with the magnetic drag models producing the
smallest shifts and the drag-free models producing the largest shifts.  The
net blueshifts for the drag-free, drag-free with added hyperdissipation, and
magnetic drag models (a) and (b) are respectively 2.07, 2.25, 1.33, and 0.99 
km~s$^{-1}$.  The shifts are as expected with the highest drag models producing
the lowest net velocity shifts.  However, given that the difference in the 
magnitude of the Doppler shifts between models only varies by 1.25 km~s$^{-1}$,
observations with RV precision of better than 1 km~s$^{-1}$ will be necessary 
to differentiate between atmospheres with and without magnetic drag.  At the 1 
km~s$^{-1}$ RV precision of the \citet{sne10} measurements, all four of our 
model atmospheres are consistent with the measured winds in HD 209458b's 
atmosphere of $2 \pm 1$ km~s$^{-1}$.

It is interesting to note that even though the drag-free model has higher 
maximum wind speeds than the drag free $+$ hyperdissipation model, the largest
Doppler shifts take place in the latter case.  This is because the extra 
hyperdissipation reduces the small-scale numerical noise observed in the 
original model and results in a more coherent flow.  Although the average wind 
speed across the terminator is the same in both models, the winds are more 
well aligned in the drag free $+$ hyperdissipation model and contribute more 
strongly to the blue shifted signal.  Although the effect is small, on the 
order of $\sim200$ m/s, this demonstrates the observational uncertainty 
associated with the choice of hyperdissipation strength in our 3-D models.

When the effects of winds are ignored and only the RV effect of rotation is
considered, a zero net Doppler shift is expected because the blueshifting of
the eastern limb of the planet should counteract the redshifting of the western
limb, resulting in Doppler broadening but no net shift.  However, 
cross-correlating the rotationally broadened spectra against the unshifted
spectra (not shown), results in a small \textit{redshift} for all of our models
of $\sim 100$ m~s$^{-1}$.  This results from the fact that the eastern 
(blueshifted) limb of the planet tends to be hotter and therefore more puffed
up than the western limb.  Absorption along the terminator on the cooler limb
therefore takes place at longitude angles where $\sin \theta$ is closer to 1.0,
which results in the rotation velocity having a larger component along the
observer's line of sight according to Equation~\ref{vlos}.  As a consequence, 
the redshifted limb plays the dominant role in the rotational broadening 
signature.  This effect, while interesting, requires measuring the rotational 
asymmetry in the transmission line profiles at a RV prevision that is an order
of magnitude higher than what has previously been obtained by \citet{sne10}.  
Even if this were possible, it would be nearly impossible to disentangle the 
rotational Doppler shifts from the Doppler shifts that result from the planet's
winds.  

\subsection{Constraints from Ingress and Egress}

Currently, measurements of Doppler shifted winds via transmission spectroscopy
require that the transmission spectra be integrated over the full transit just
to attain the signal-to-noise necessary for making a 1 km~s$^{-1}$ measurement.
However, future observations that can measure Doppler shifts as a function of
orbital phase throughout transit will be a very powerful tool for resolving 
the spatial structure of the winds across a planet's terminator.   These types
of measurements will likely necessitate next generation ground-based 30-m class
telescopes to attain high signal-to-noise over shorter exposure times.  Of
particular interest is the measurement of Doppler shifts during transit ingress
and egress when only one limb of the planet is in front of the star, which will
allow for a straightforward mapping of the eastward and westward flows on 
either limb of the planet.  

We calculate Doppler shifts as a function of orbital phase throughout transit 
for each of the four model atmospheres in Figure~\ref{phase_shift}.  We 
separately show the cumulative effects of rotation, winds, and orbital motion.
Generally, the orbital motion is the dominant effect on the Doppler-shifted
signal.  When that motion is subtracted off, as in \citet{sne10}, the effects 
of winds become apparent.  

Some details of the calculation resulting in Figure~\ref{phase_shift} are as
follows.  Throughout the analysis we use a single snapshot from each model, 
safely assuming that the amount of temporal variation is negligible and its 
effect is small compared to our other simplifying assumptions.  We account for 
rotation of the planet away from phase $\varphi = 0$ by shifting the 
longitudes in our model by an angle equivalent to the orbital phase.  This step
is important because the planet rotates by an angle of 16$^{\circ}$ over the 
duration of the transit relative to an observer on Earth, whereas the frame of 
reference of the model atmosphere calculations always has the 0$^{\circ}$ line 
of longitude at planet's substellar point.  The planet's rotation relative to 
the observer has only a small effect on the calculated velocities, but the 
effects of the slow eastward motion of both the hottest and coldest points on 
the planet does induce small velocity shifts throughout transit.  For 
simplicity, we assume that the planet transits exactly across the middle of the
star at an orbital inclination of 90$^{\circ}$.  We do not include any effects 
of limb darkening.  We also ignore the geometric effects of curvature on the 
limb of the star, in effect treating the edge of the star as a straight line, 
tangential to the orbital direction of the planet.  These last three effects 
should minimally impact the results shown in Figure~\ref{phase_shift}, but the 
qualitative results are still instructive. 

During full transit, between the 2nd and 3rd contact points when the planet is 
completely in front of the star, the wind and rotation RVs are fairly constant,
despite a very small effect from the rotation of the planet relative to the
observer, which can cause shifts of up to 200 m~s$^{-1}$.  In contrast, the 
orbital RV signature varies considerably during transit as it sweeps out the 
$\varphi = 0$ region of a sine curve with peak amplitude 146 km~s$^{-1}$, 
equivalent to the orbital speed of the planet on its circular orbit.  

However, during transit ingress and egress, between the 1st and 2nd, and 3rd 
and 4th contact points respectively, the induced Doppler shifts from winds and
rotation both vary considerably as first one limb of the planet and then the 
other appears/disappears in front of the stellar disk.  The effect of the 
planet's rotation is to blueshift the eastward limb and redshift the westward
limb of the planet, thus creating an anomalous redshift during ingress and
blueshift during egress.  Since the rotation is symmetric, the rotational RV
signature is also approximately (anti-)symmetric about the center of transit.  
The magnitude of the Doppler shift due to rotation alone increases from 
$\sim 0$ km~s$^{-1}$ during the phase of full transit to $\sim 2$ km~s$^{-1}$ 
at the very beginning of ingress and end of egress, which is equivalent to the 
rotational speed of the planet, as seen in the bottom panel of 
Figure~\ref{phase_shift}.  

Winds display a similar behavior as rotation during ingress and egress where 
the eastward limb tends to be more strongly blueshifted than the westward limb.
This effect varies from model to model and also becomes stronger with depth, as
shown in Figure~\ref{fig:dshift}.  In general, most of the terminator is 
predicted to be blueshifted at mbar pressures.  However, the models with 
magnetic drag tend to have more uniform velocities, whereas the drag-free 
models show a stronger tendency towards blueshifted winds at the eastern 
terminator and redshifted winds at the western terminator.  In terms of the 
ingress and egress velocities, the canonical drag-free model gives the largest 
differences between the Doppler shifts from winds at the start and finish of 
transit.  At the very beginning of ingress the drag-free model gives an 
additional 1.6 km~s$^{-1}$ blueshift on top of the 2 km~s$^{-1}$ redshift from 
the planet's rotation (for a net redshift of 0.4 km~s$^{-1}$).  At the very
end of egress this same model gives an additional 3.3 km~s$^{-1}$ blueshift on 
top of the 2 km~s$^{-1}$ rotational blueshift.  This results in an 
overall 1.7 km~s$^{-1}$ measured difference between the line-of-sight 
wind speeds for opposing limbs of the planet.  In contrast, the winds in the 
models with magnetic drag each only produce a 0.2 km~s$^{-1}$ difference in the
Doppler shifts between the leading and trailing limbs.  If this effect can be 
measured, it allows for another constraint to be placed on the amount of drag 
present in a hot Jupiter atmosphere.  It serves pointing out that correctly 
measuring the Doppler shifts that arise exclusively from winds requires that 
both the orbital and rotational effects be subtracted off, which necessitates 
an assumption of tidal locking along with knowledge of the orbital speed.  

Measuring velocity shifts during ingress and egress faces two key challenges.
The first, which has already been mentioned, is that high signal-to-noise
spectra must be obtained using fairly short exposures since transit 
ingress/egress may only each last for 10-30 min in total.  A second major 
challenge to obtaining high signal-to-noise during ingress/egress is that the 
excess absorption from the planet's atmosphere that creates the transmission 
spectrum decreases in magnitude as the planet moves off of the stellar disk.  
The amount of excess absorption scales linearly with the cross sectional area 
of atmosphere that is in front of the star at a given time, which quickly 
reduces to zero at the beginning and end of transit.  With almost no photons 
passing directly through the planet's atmosphere near 1st and 4th contact, it 
will be impossible to measure any meaningful Doppler shifts, even if the winds 
at the edge of the planet's limb are particularly strong.  Doppler shift 
measurements during ingress/egress should still be possible when a larger 
fraction of the planet is in front of its host star, but the signal-to-noise 
requirements will almost certainly require next generation telescopes and 
instrumentation.  

\subsection{Constraints from Individual Spectral Lines}

Another interesting constraint can be placed on the spatial structure of the
winds by measuring Doppler shifts from individual spectral lines.  Stronger
spectral lines in transmission result from the planet becoming optically thick 
at larger radii, so these lines originate higher in the atmosphere.  Therefore,
by measuring velocity shifts in individual spectral lines as a function of 
their peak strength, a measurement can be made of the vertical structure of
the exoplanet winds.  If stronger lines produce larger Doppler shifts, then 
the winds higher in the atmosphere are stronger, whereas weaker Doppler shifts
at altitude imply that wind speeds fall off as a function of height.  

Our drag-free atmosphere models tend to produce higher line-of-sight wind
speeds at altitude, whereas the magnetic drag models have very little wind 
shear or even opposite wind shear in the case of magnetic drag model (b) (see 
Figure~\ref{fig:dshift}).  To test the extent to which this could be an 
observable effect, we determine Doppler shifts for 134 individual spectral 
lines relative to their rest wavelengths.  The results for each of the four
atmosphere models are shown in Figure~\ref{line_strength}.  As predicted, the
drag-free models show increasing velocity (blue) shifts as a function of 
increasing line strength, whereas magnetic drag model (a) shows almost no 
RV variation as a function of height, and model (b) produces a slight trend
towards decreasing velocity shifts at altitude.  The line shifts were all 
calculated by measuring the shift in the wavelength of peak intensity from
the unshifted model to the fully Doppler shifted model.  For this reason, 
velocity shifts are only measured in integer multiples of the velocity
resolution of our $R = 10^6$ spectra, resulting in the vertical alignment 
seen amongst groups of data points in Figure~\ref{line_strength}.  

Linear fits to the line shift data for each model are also shown in 
Figure~\ref{line_strength} to guide the eye, as there is considerable scatter 
in the measurements.  The 134 lines sampled include spectral lines from CO, 
H$_2$O, and CH$_4$, which likely increases the scatter in the line strength 
vs.~velocity relationship.  This is particularly true for CO and CH$_4$ because
lines from these two molecules tend to originate from different regions of the
atmosphere -- CO from regions of hotter gas, likely weighted towards the 
eastern terminator and CH$_4$ from regions of colder gas on the western 
terminator.  Measuring the line strengths for only the 76 first-overtone lines
of CO (not shown) results in a tighter linear relationship for line strength 
vs.~velocity.  However, this particular set of lines samples a far smaller 
range of altitude in the atmosphere, since all of the CO spectral lines are 
quite strong and therefore originate from regions of high altitude and low 
pressure.

Measuring Doppler shifts of individual spectra lines in transmission is another
effect that will almost certainly require next generation instrumentation to 
realize, due to the very high signal-to-noise requirements.  The only 
measurements of Doppler shifts in transmission to date \citep{sne10} required
the full spectrum from 2291 to 2349 nm to detect a Doppler shift at km~s$^{-1}$
precision.  Individual spectral lines were not present in their spectra at
close to sufficient signal-to-noise to measure RV shifts, even when 
integrating over a full transit.  

\section{Summary / Conclusion \label{concl}}

We have coupled together an existing 3-D atmospheric dynamics model with a
transmission spectroscopy radiative transfer code to study the effects of 
Doppler shifted winds (along with rotation and orbital motion) on the 
transmission spectra of hot Jupiter exoplanets.  We find that day-to-night
winds at altitudes of $\lesssim$ 1 mbar can produce significant blueshifts
in hot Jupiter transmission spectra at the level of 1-2 km~s$^{-1}$.  Also, the
combined effects of winds and the planet's rotation lead to considerable 
Doppler broadening of the spectral lines beyond what is predicted if no Doppler
shifts are included.  These effects all become important for transmission 
spectra taken at high spectral resolution of $R \gtrsim 10^5$.  We have 
explored different prescriptions for magnetic drag in the atmospheres of hot 
Jupiters, and we find that the models with the largest amount of magnetic drag 
produce the slowest wind speeds and therefore the smallest Doppler (blue) 
shifts, while the models with no magnetic drag produce the largest velocity 
shifts.  In our modeling, the magnetic drag models produce net blueshifts of 
$\sim$1 km~s$^{-1}$, whereas the models with no drag produce blueshifts of 
$\sim$2 km~s$^{-1}$.  Ultimately differentiating between different drag 
prescriptions through observations of hot Jupiter transmission spectra will 
therefore require RV precision of much higher than 1 km~s$^{-1}$.  In the 
meantime, the current measurement of a $2 \pm 1$ km/s blueshift in HD 
209458b's transmission spectrum by \citet{sne10} remains consistent with 
models both with and without magnetic drag.  

Still, the results of our modeling give considerable credence to the
interpretation that the blueshift observed in HD 209458b's transmission 
spectrum results from high altitude winds, as our models produce Doppler shifts
of exactly that size.  An intriguing alternate interpretation of the 
\citet{sne10} observed blueshift has been proposed by \citet{mon11}, who 
show that a small eccentricity in the orbit of HD 209458b can also produce a
net blueshift in the planet's transmission spectrum from orbital motion alone. 
The eccentricity of HD 209458b's orbit is strongly constrained and consistent 
with zero \citep{win05, dem05}.  However, using the 3-$\sigma$ upper limits on 
the planet's eccentricity, \citet{mon11} find that a blueshift of up to 1 
km~s$^{-1}$ could result from a non-circular orbit, which is consistent with 
the \citet{sne10} observed velocity shift within its error bars.  Ultimately, 
better constraints on the orbital eccentricity of HD 209458b will be needed to
resolve whether winds or a non-circular orbit are the cause for the observed
blueshift in the planet's spectrum.  In the meantime, a probable very low 
eccentricity for HD 209458b does require that winds play the dominant role in 
producing km~s$^{-1}$ blueshifts in the planet's transmission spectrum.  

Even more ambitious observations aimed at spatially 
resolving exoplanet winds by observing Doppler shifts for individual spectral 
lines and Doppler shifts as a function of transit phase will likely remain 
beyond the reach of current observational facilities.  Future instrumentation 
-- high resolution spectrographs with $R \gtrsim 10^5$ on next generation 30-m
class telescopes -- may be able to measure some of these effects.  A 
combination of vertical wind shear measurements obtained from the Doppler
shifts of individual spectral lines along with ingress and egress measurements
of wind speeds on opposing limbs of the planet can produce a 3-D ``map'' of
winds along an exoplanet's terminator.  This is of particular interest for
tidally locked hot Jupiters, since the winds at the terminator are intimately
tied to the day-to-night heat flow and therefore to the global energy budget
of the planet.

We caution that the form for magnetic drag used in these 3-D models is 
simplistic and a more realistic treatment could result in a different flow 
pattern, which would alter the details of the predicted ingress/egress and 
vertical shear measurements.  However, with these models we have been able to 
demonstrate the types of observations that would constrain the 3-D atmospheric 
circulation.  Regardless of the exact form for the magnetic drag, what 
\textit{is} a fairly robust result is that the drag should work to reduce wind 
speeds, resulting in smaller Doppler shifts in the observed spectra.

In a particularly interesting proof of concept, \citet{hed11} observed the 
transmission spectrum of Venus during its 2004 transit of the Sun and 
discovered Doppler shifted spectral line profiles.  While many differences 
exist between Venus and hot Jupiters, the very slow (retrograde) rotation of 
Venus means that it has very long days ($\sim$0.5 Venus years), compared to hot
Jupiters' permanent days, resulting in some similarities between their 
circulation regimes.  A full analysis of Venus' Doppler shifted line profiles 
is beyond the scope of the \citet{hed11} paper, but the authors comment that 
the observed Doppler shifts likely result from a combination of super-rotation 
of Venus' mesosphere along with day-to-night flow -- similar to what we predict
for hot Jupiter atmospheres but at lower velocities.  Using spatially resolved 
spectra, \citet{hed11} were furthermore able to separately measure the wind 
speeds on both the leading and trailing limbs of the planet and also to 
directly measure the vertical wind shear.  The high resolution and high 
signal-to-noise transmission spectra obtained by \citet{hed11} show the power 
of using transmission spectroscopy to constrain the wind patterns in the upper 
atmospheres of extrasolar planets.    

Ultimately, measurements of winds in hot Jupiter atmospheres will be able to 
constrain the circulation and flow patterns in an interesting new regime.  The
flow patterns that are expected to result from the tidal locking and very hot
dayside temperatures of these planets require observational confirmation, and
we have laid out some of the methods for modeling and interpreting those 
observations throughout this paper. The observation of a blueshifted 
transmission spectrum for HD 209458b by \citet{sne10} was an exciting first 
step towards observationally constraining the wind patterns on hot Jupiters.  
Similar observations of a larger number of hot Jupiters will be able to confirm
whether the blueshifted line profiles resulting from high altitude day-to-night
winds are ubiquitous or vary considerably from planet to planet.  This will 
help to better understand diversity in hot Jupiter circulation (as indicated by
differences in planets' thermal phase curves), especially as is related to the 
effect of magnetic drag, which will depend on a given planet's atmospheric 
temperatures and magnetic field strength.

Currently, the available instrumentation is a limiting factor in measuring and 
constraining the wind patterns on hot Jupiters.  Next generation 
instrumentation will allow for higher precision and more detailed measurements 
to be taken of a larger number of hot Jupiter atmospheres.  Combining these 
observations with the results from theoretical modeling will provide powerful
and unique constraints on day-to-night flow on hot Jupiters.

\begin{figure}
\begin{center}
\plotone{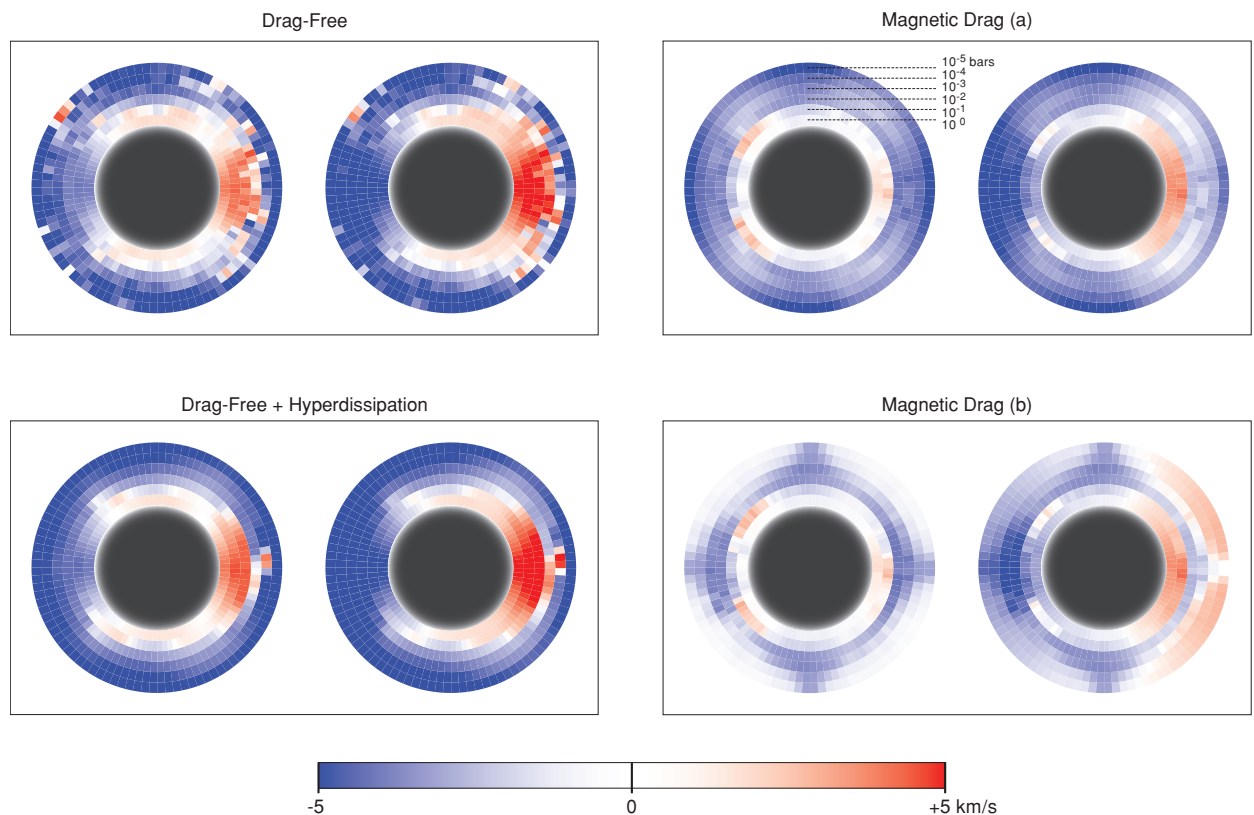}
\end{center}
\caption{Line-of-sight velocities at the terminator for the various models.  
        The left-hand image in each pair shows the line-of-sight wind speeds in
	the frame of the planet, whereas the right-hand image is in the frame 
	of the observer and includes the effect of the planet's rotation.  All 
	snapshots are taken at orbital phase $\varphi = 0$ where the orbital 
	motion has a null effect on the line-of-sight velocity.  The form and 
	strength of hyperdissipation and/or magnetic drag changes the 
	line-of-sight wind speeds as well as the pressures at which they are 
	the strongest.  Rotation always blueshifts the east terminator and 
	redshifts the west, at a magnitude of $\sim$2 km s$^{-1}$ at the
	equator.  Line-of-sight velocities over 5 km s$^{-1}$ are shown in the 
	darkest shade of red/blue.} 
\label{fig:dshift}
\end{figure}

\begin{figure}
\begin{center}
\includegraphics[scale=0.72, angle=90]{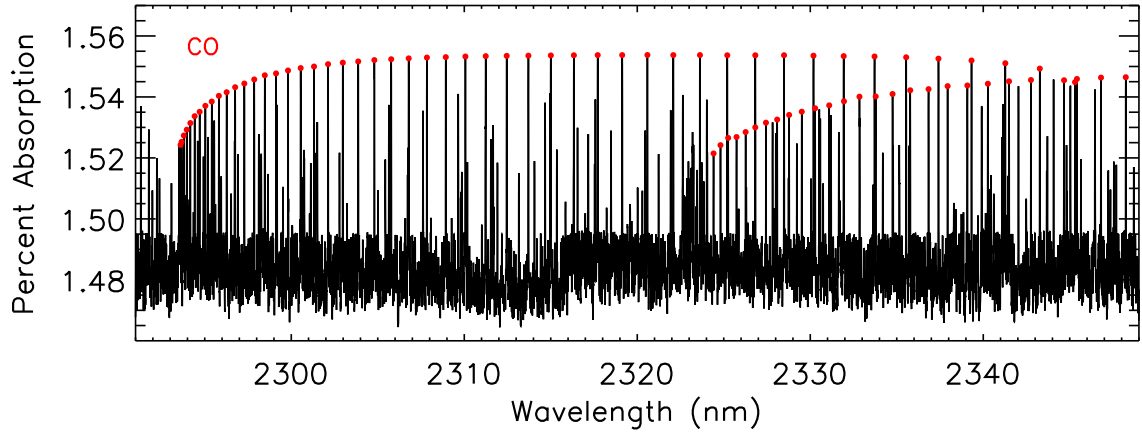}
\end{center}
\caption{Unshifted transmission spectrum calculated from 2291 to 2349 nm.  The
        y-axis plots the transit depth in percent as a function of wavelength.
	The strong regularly-spaced spectral lines are the 2.3 $\mu$m first 
	overtone band of CO.  (The $\nu = 2-0$ and $\nu = 3-1$ bands of CO are
	indicated in red.)  The remainder of the absorption lines in this
	region of the spectrum are mostly from H$_2$O and CH$_4$, with a much 
	smaller number of weak lines from NH$_3$.
        \label{transmission_full}}
\end{figure}

\begin{figure}
\begin{center}
\includegraphics[scale=0.68, angle=90]{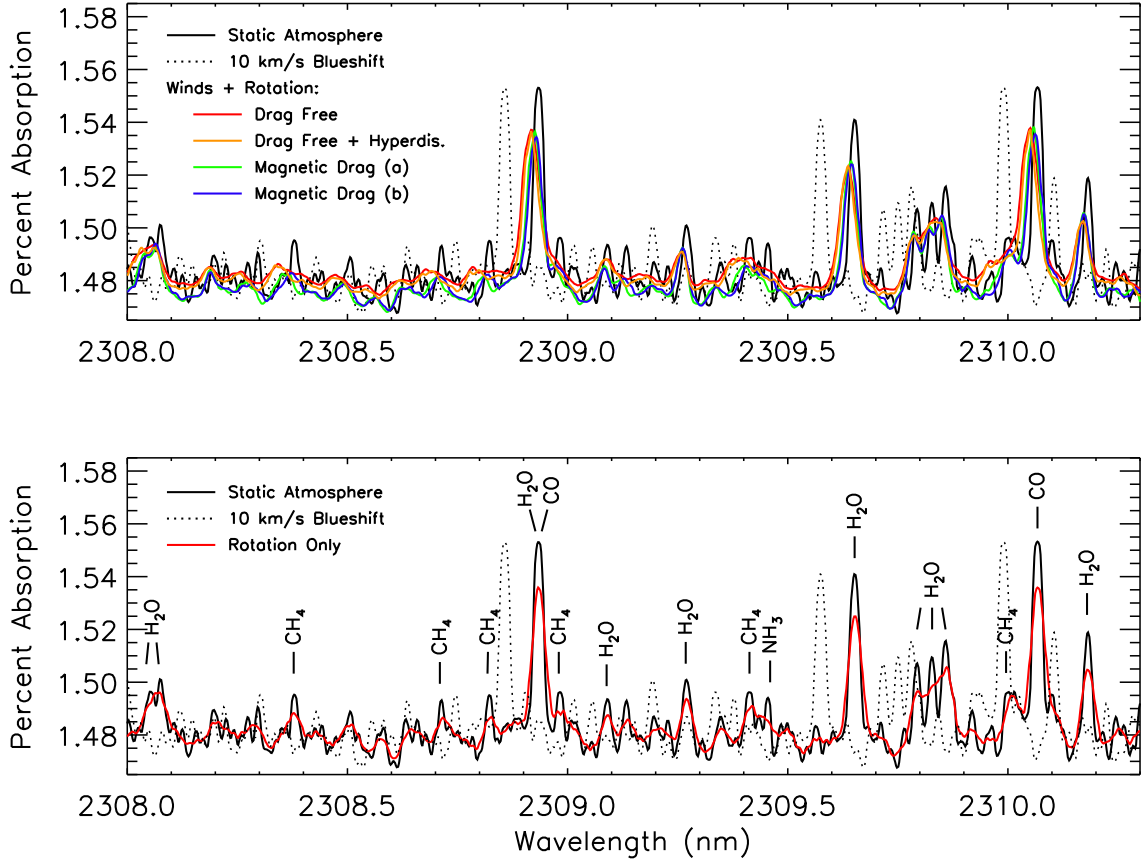}
\end{center}
\caption{Snapshots of a 2.3 nm section of the Doppler-shifted transmission 
        spectra taken at the center of transit ($\varphi = 0$).  \textit{Top}: 
	Doppler shifted spectra including the effects of both winds and 
	rotation.  All models reveal a net blueshift that varies from 1-2 
	km~s$^{-1}$ depending on the drag prescription.  The unshifted 
	transmission spectrum for the drag-free case is shown 
	(\textit{black solid}), as is the same spectrum with a constant 10 
	km~s$^{-1}$ offset for reference (\textit{black dotted}). 
	\textit{Bottom}: Same as above but only the effect of rotation is 
	included, which broadens the spectrum but does not induce a significant
	RV shift.  Only the drag-free case is shown, as all four models produce
	almost identical rotational signatures.  \label{spectra_snapshot}}
\end{figure}

\begin{figure}
\plotone{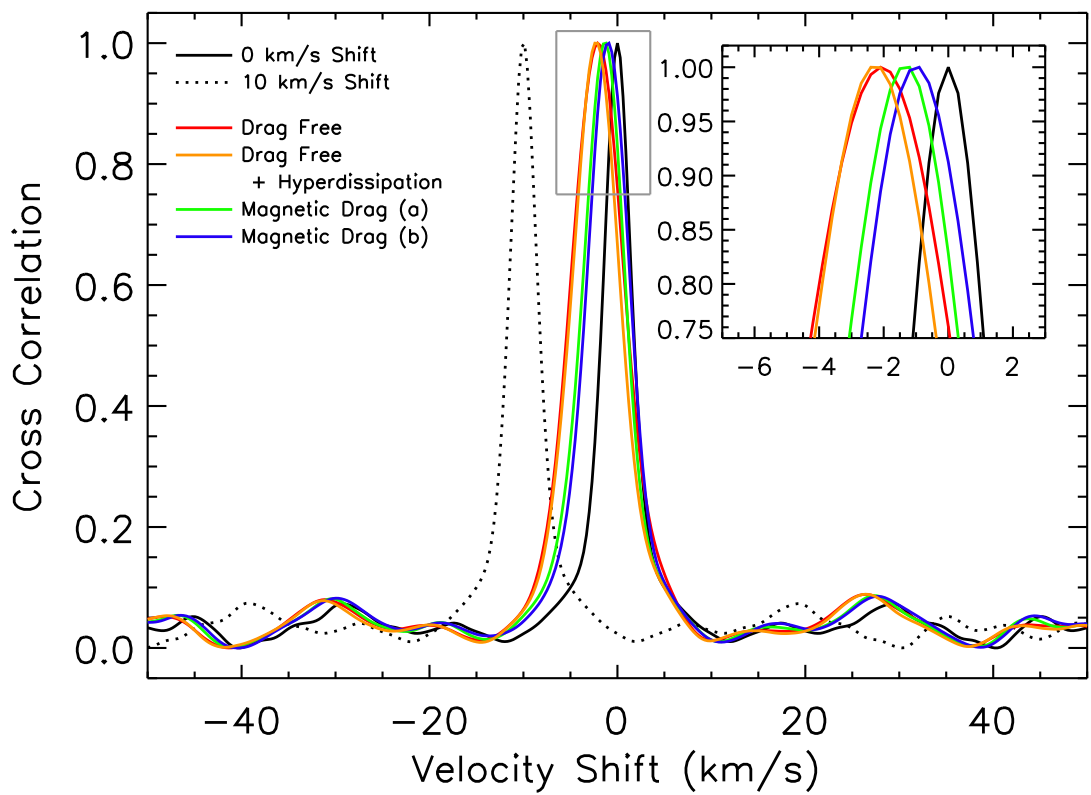}
\caption{Cross correlation functions for each of the four atmosphere models
        taken at the center of transit against the unshifted spectra.  The 
	cross correlation function for the unshifted drag-free spectrum against
	itself (\textit{black}) and with a 10 km~s$^{-1}$ blueshift 
	(\textit{dotted black}) are shown for reference.  The strongest RV 
	shifts belong to the models with the weakest drag prescriptions and 
	vice versa.  \label{crscor}}
\end{figure}

\begin{figure}
\begin{center}
\includegraphics[scale=0.97]{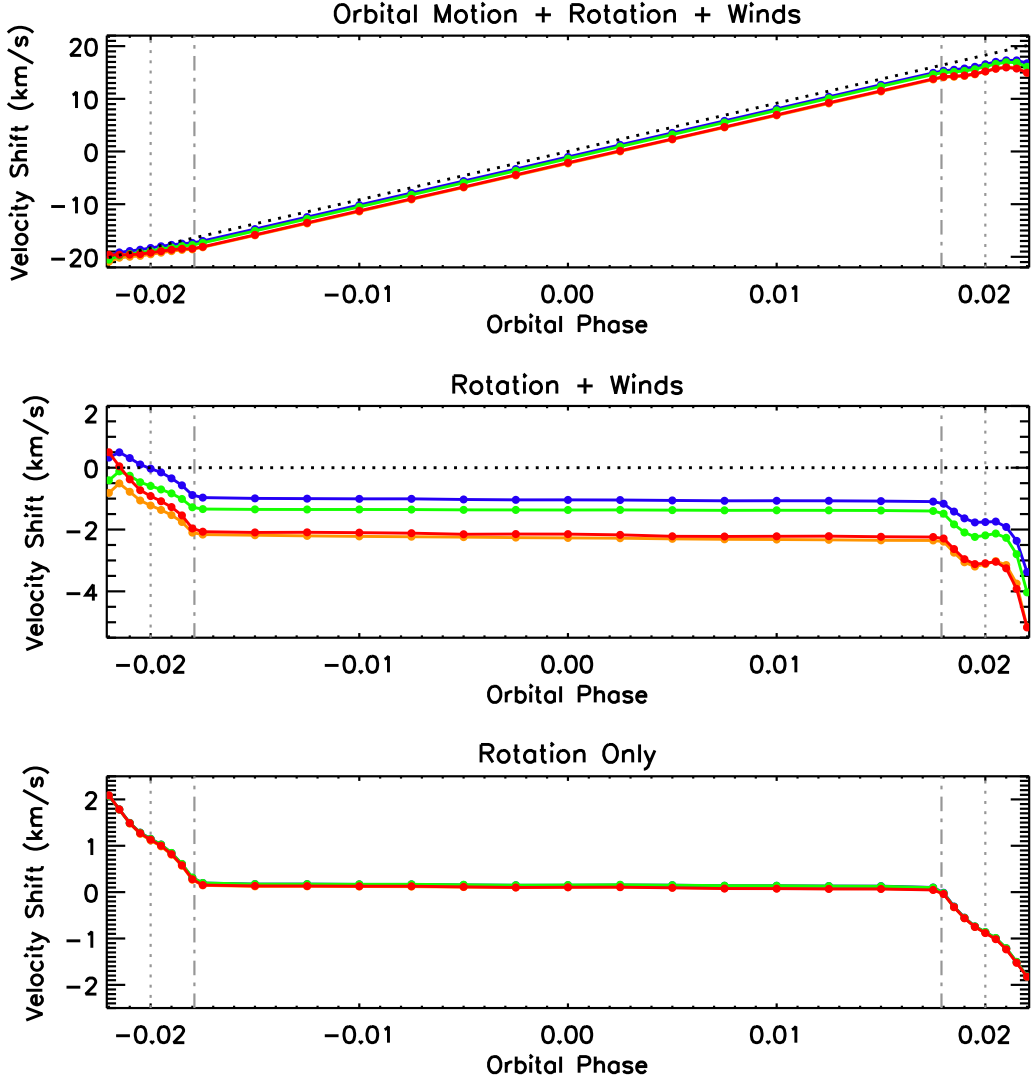}
\end{center}
\caption{Velocity shifts as a function of orbital phase.  \textit{Top}: The
        cumulative RV effects of orbital motion, winds, and rotation are shown.
	The dotted black line indicates the Doppler shift that would be 
	expected if orbital motion was the only contributor to the RV signal.  
	\textit{Middle}: The cumulative RV effects of winds and rotation.  The
	dotted black curve from the top panel has been subtracted off (now 
	shown as a horizontal line at 0 km~s$^{-1}$), revealing a blueshifted
	offset for all four models that results from day-to-night winds.  
	\textit{Bottom}:  The RV effects of rotation alone.  The left-hand and 
	right-hand boundaries of the plots are at the 1st and 4th contacts of 
	transit.  Gray vertical dot-dashed lines indicate the 2nd and 3rd 
	contacts of transit, and dotted lines indicate the ``1.5th'' and 
	``3.5th'' contact points where one limb of the planet each is on and 
	off of the star.  All effects of limb darkening have been ignored.  
	\label{phase_shift}}
\end{figure}

\begin{figure}
\plotone{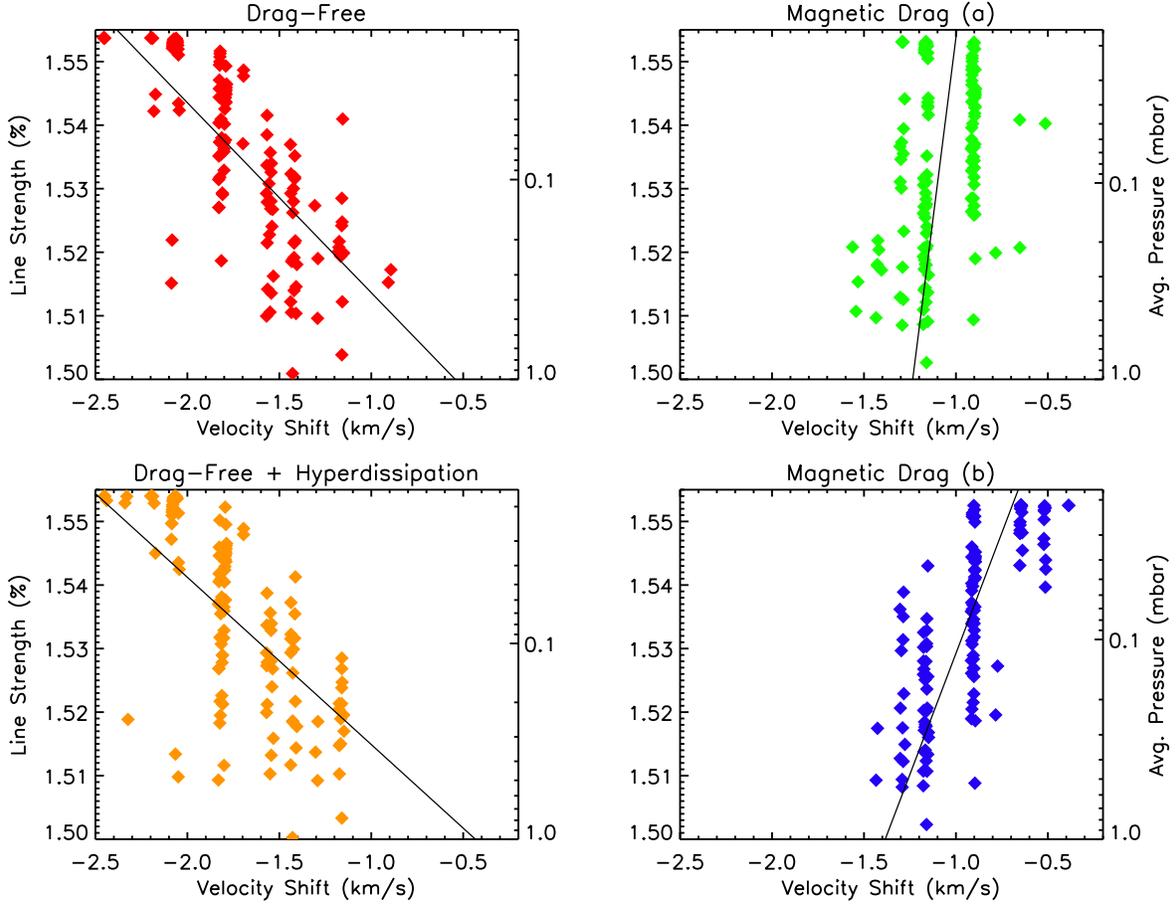}
\caption{Strength of individual spectral lines (i.e.~their peak transit depth 
        in percent) vs.~Doppler shift.  The stronger lines originate from 
	higher in the atmosphere.  Shifts for 134 individual lines are plotted 
	with a linear fit overlaid.  The average atmospheric pressure 
	corresponding to a given line strength is also indicated.  The vertical
	alignment (striping) in the data is due to the velocity resolution of 
	our $R=10^6$ spectra.  
        \label{line_strength}}
\end{figure}

\acknowledgements 
E.~M.-R.~K and E.~R. were supported by contracts with the California Institute 
of Technology, funded by NASA through the Sagan Fellowship Program.  We thank
Paul A. Kempton for his design and collaboration on Figure~\ref{fig:dshift}.  
We thank Kristen Menou for encouraging this work.  Both E.~M.-R.~K and E.~R. 
also credit the AstroWin 2011 workshop at the University of Florida for hosting
us and encouraging the collaboration the resulted in this paper.

\bibliography{ms}

\end{document}